\documentclass[final,5p,times,twocolumn]{elsarticle}
\usepackage{graphicx,color,epsfig,rotate,amssymb}
\usepackage{hyperref}

\newcommand{\Eqn}[1] {Eqn.~(\ref{#1})}

\newcommand{\Fig}[1]{Fig.~\ref{#1}}

\journal{Solid State Communications}

\begin{document}

\begin{frontmatter}

\title{Orbital Magnetic Field Driven Metal-Insulator Transition in Spinless Extended Falicov-Kimball Model on A Triangular Lattice}

\author{Umesh K. Yadav}
\address{Department of Physics, Lovely Professional University, Phagwara - 144411, Punjab, India}

\begin{abstract}
Ground state properties of spinless, extended Falicov-Kimball model (FKM) on a finite size triangular lattice with orbital magnetic field normal to the lattice are studied using numerical diagonalization and Monte-Carlo simulation methods. We show that the ground state configurations of localized electrons strongly depend on the magnetic field. Magnetic field induces a metal to insulator transition accompanied by segregated phase to an ordered regular phase except at density $n_f = 1/2$ of localized electrons. It is proposed that magnetic field can be used as a new tool to produce segregated phase which was otherwise accessible only either with correlated hopping or with large on-site interactions.
\end{abstract}


\end{frontmatter}

\section{Introduction}
Electrons moving in a periodic potential under the influence of external magnetic field is an extensively studied problem in low dimensional systems, giving rise to many novel phenomenon like Quantum Hall effect~\cite{Hall1,Hall2} and Hofstadter butterfly~\cite{Hofst} structure in the non-interacting limit. Electron correlations play an important role to govern the properties of the systems in low dimensions. In the limit of finite electron correlations very little results are known because of complex nature of problem. 

Systems like cobaltates~\cite{qian06,tera97,tekada03}, $GdI_{2}$~\cite{tara08} and its doped variant $GdI_{2}H_{x}$~\cite{tulika06}, $NaTiO_{2}$~\cite{clarke98,pen97,khom05}, MgV$_{2}$O$_{4}$~\cite{rmn13} etc. have attracted great interest as they exhibit a number of remarkable cooperative phenomena such as valence and metal-insulator transition, charge, orbital and magnetic order, excitonic instability and possible non-fermi liquid states~\cite{tara08}. In these systems ordering is governed by interplay between kinetic and interaction energies of electrons and underlying lattices. These are layered, triangular lattice systems and are characterized by the presence of localized ($f$-) and itinerant ($d$-) electrons. The geometrical frustration from underlying triangular lattice coupled with strong quantum fluctuations give rise to a huge degeneracy at low temperatures resulting in competing ground states close by in energy. Therefore, for these systems one would expect a fairly complex ground state phase diagram and the presence of soft local modes strongly coupled with the itinerant electrons. 

It has proposed that these systems may very well be described by different variants of the Falicov-Kimball model (FKM)~\cite{tara08,tulika06} on a triangular lattice. The FKM was introduced to study the metal-insulator transitions in the rare-earth and transition-metal compounds~\cite{fkm69,fkm70}. The model has also been used to describe a variety of many-body phenomenon such as tendency of formation of charge and spin density wave, mixed valence, electronic ferroelectricity and crystallization in binary alloys~\cite{lemanski05,farkov02,taraph01}.

Recently ground state properties of spinless FKM on a triangular lattice with correlated hopping ($t^{\prime}$) are reported~\cite{umesh1}. The correlated hopping term was taken into account due to fact that in some rare earth compounds (specially the mixed-valence compounds), the rare earth ions with two different ionic configurations $f^{n}$ and $f^{n-1}$ have different ionic radii. The contraction of $d$-orbitals in ions having $f^{n-1}$ configuration, for reduced screening of nuclear charge, leads to increased localization of $d$-electrons. Hence the $d$-orbital overlap between nearest neighbors depends on local $f$-electron configurations of neighboring ions, resulting in a correlated hopping ($t^{\prime}$) of $d$-electrons. In this study several charge ordered states of varying periodicities and phase segregated configurations was reported for different parameters e.g. filling, on-site interaction between $d$- and $f$-electrons ($U$) and correlated hopping $t^\prime$. In presence of correlated hopping phase segregation was seen even at smaller $U$-values. A rich phase diagram of localized electrons was reported due to combined effects of correlation and frustration on a triangular lattice. 

Many numerical and exact calculations are available for different extensions of Falicov-Kimball model on bipartite and non-bipatatite lattices in the absence of magnetic filed taking into account interactions between itinearant and localized electrons~\cite{lemanski05,farkov02,umesh2,umesh3,umesh4,umesh5}. But there is only few results are available for FKM with finite magnetic field~\cite{grub,sq_lat,taraph01}. 

In a very recent study ground state properties of FKM with orbital magnetic field perpendicular to the triangular lattice are reported for $U = 0$ and for large  $U$ cases~\cite{taraph02}. Combined effects of magnetic field and on-site interactions on ground state properties are reported by studying energy spectrum of itinerant electrons. At all chosen values of $U$, it is shown that magnetic field affects the energy spectrum. They have reported an ordered ground state configurations and found that at certain values of flux, gap at Fermi-level in energy spectrum decreases.  Also, at all values of flux a finite gap is seen in the energy spectrum for finite electron correlations.  

There are no rigorous results are available for the FKM on triangular lattice with finite magnetic flux in all range of on-site interaction $U$. Specifically it would be very interesting to see the effects of magnetic field on the ground state configurations of localized electrons in the limit of $U \sim t$. In the presence of magnetic field and on-site interaction there will be competition between different length scales namely cyclotron radius, inter-atomic distance and type of patterns in the ground state configuration of $f$-electrons. Exploring the role of magnetic field on the ground state configurations in the presence of correlated hopping $t^{\prime}$ will also be quite interesting. 

Therefore, in the present paper we have explored the spinless extended Falicov-Kimball model on a triangular lattice with finite magnetic flux in each traingle. We have studied the ground state configuarations and their nature (metallic/insulating) for different range of parameters like $U$, $t^{\prime}$ and filling of electrons. These results on FKM with magnetic field are also relevant in the context of cold atomic systems. There are recent proposals that FK Hamiltonian can be seen in the lab using optical lattices loaded with mixture of light and heavy atoms for both bosonic and fermionic cases~\cite{fkm_opt_lat01,fkm_opt_lat02,fkm_opt_lat03}.                        

\begin{figure}
\begin{center}
\includegraphics[width=9.0cm]{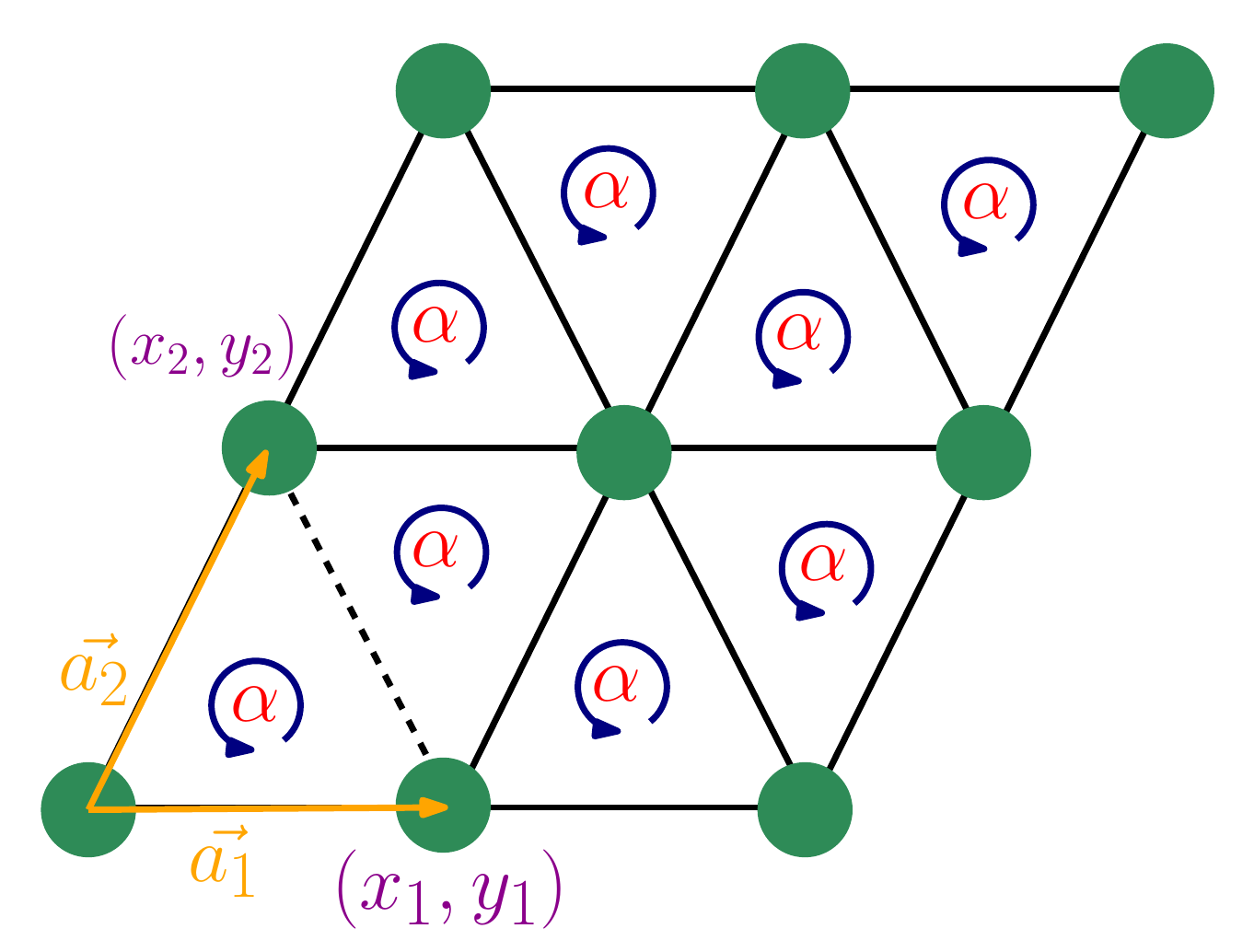}
\end{center}
\caption{(Color online) Schematic plot of a triangular lattice of size $\left(3 \times 3 \right)$ with an uniform magnetic flux $\alpha$ in each triangle is shown. A particular bond on a triangle is shown by dotted line with coordinates $(x_1,y_1)$ and $(x_2,y_2)$. Direct primitive lattice vectors on a triangular lattice are shown in orange color and given by $\vec{a_{1}} = a(1, 0)$ and $\vec{a_{2}} = a(\frac{1}{2}, \frac{\sqrt{3}}{2})$, where a is the lattice constant. These primitive lattice vectors are used to generate the coordinates of bonds on each triangle. Arrow shown in blue color in each triangle represents direction of hopping of itinerant electrons in a triangle.}
\label{fig:triang_lat}
\end{figure}

The FKM Hamiltonian with finite magnetic flux in each triangle is given as,
\begin{eqnarray}\label{eqn:Ham01}
H  =\,\sum\limits_{\langle ij\rangle} \,\bigg[ \big\{ -t_{ij} + t_{ij}^{\prime} \left(f_{i}^{\dagger} f_{i} + \,f_{j}^{\dagger} f_{j}\right) \big\} 
\nonumber \\
\exp{\left( \pm\,\frac{\mathit{ie}}{\hbar\mathit{c}} \int\limits_{\vec{R_i}}^{\vec{R_j}} \bf{A}(\vec{r}) \cdot d\vec{r} \right)} 
\,-\mu\delta_{ij}  \bigg] \,d^{\dagger}_{i}d_{j} + 
\,U\sum_{i}f^{\dagger}_{i}f_{i}\,d^{\dagger}_{i}d_{i}
\nonumber \\
+\,E_{f}\sum\limits_{i}f^{\dagger}_{i}f_{i}
\end{eqnarray}
\noindent here $\langle ij\rangle$ denotes the nearest neighboring ($NN$) lattice sites $i$ and $j$. The $d^{\dagger}_{i},  d_{i}\,(f^{\dagger}_{i},\,f_{i})$ are, respectively, the creation and annihilation operators for $d$- ($f$-) electrons at site $i$. First term is the kinetic energy of the $d$- electrons in the presence of orbital magnetic field, induced by vector potential $\bf{A}(\vec{r})$, affecting the hopping of $d$-electrons. Here $t$ is quantum mechanical hopping amplitude and $t^\prime$ is correlated hopping of $d$-electrons depending on the occupation of nearest-neighboring sites of $f$-electrons~\cite{umesh1}. The $\mu$ is chemical potential. Second term is on-site interaction between $d$- and $f$-electrons. The last term is the dispersionless energy level $E_f$ of $f$-electrons.

\begin{figure}
\begin{center}
\includegraphics[width=9.0cm]{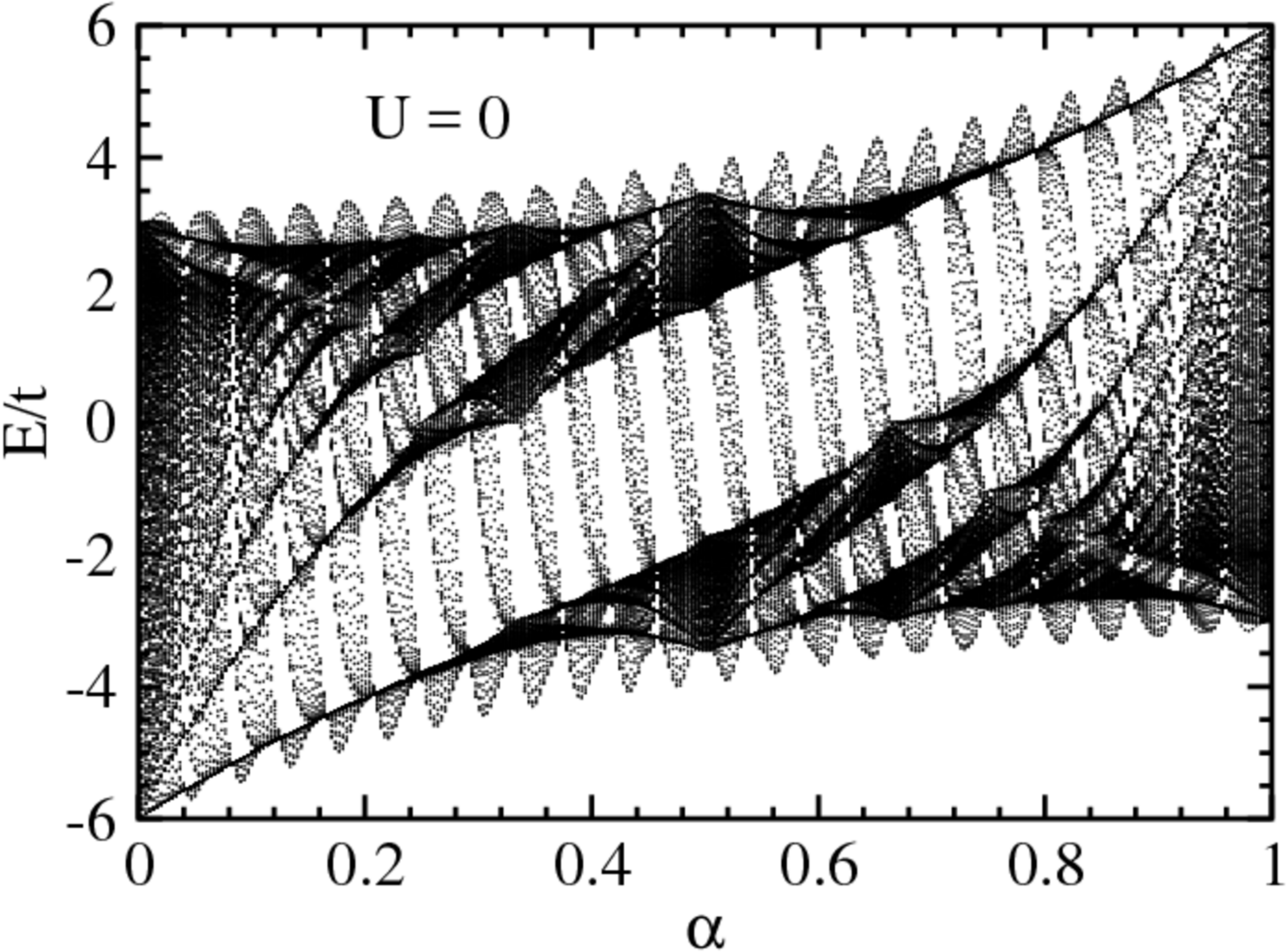}
\end{center}
\caption{Eigen spectrum of itinerant electrons as a function of magnetic flux $\alpha$ (= $\frac{\phi}{\phi_0} = \frac{\sqrt{3} a^{2}}{2} \bf{B}$) per unit cell on a triangular lattice of size ($L =24$) at $U =0$ with periodic boundary condition.}
\label{fig:hfsdtru0}
\end{figure}

\section{Methodology}

Hamiltonian $H$ (\Eqn{eqn:Ham01}), preserves the states of $f$-electrons, i.e. $d$- electrons traveling through the lattice does not  change occupation numbers of $f$-electrons. Therefore, local $f$-elctron occupation number $\hat{n}_{f\,i} =f_{i}^{\dagger}f_{i}$ is invariant and $\big[ \hat{n}_{f\,i},H \big] = 0$ for all $i$. This also shows that $\omega_{i}=f_{i}^{\dagger}f_{i}$ is a good quantum number taking values only $1$ or $0$ according to whether the site $i$ is occupied or unoccupied by $f$-electron, respectively. Following the local conservation of $f$-electron occupation and for magnetic field perpendicular to the plane of a triangle~\cite{orbital} (with Landau gauge, choosing the vector potenial as $\bf{A(\vec{r})} = \bf{B} \left(0, x, 0 \right)$~\cite{phase}),  $H$ can be written as,
\begin{eqnarray}\label{eqn:Ham02}
H  =\,\sum\limits_{\langle ij\rangle} h_{ij} d_{i}^{\dagger} d_{j} + \,E_{f}\sum\limits_{i}\omega_{i}
\end{eqnarray}
\noindent with, \begin{eqnarray}\label{eqn:ham}
h_{ij}=\big[ \big\{ -t_{ij} + t_{ij}^{\prime} (\omega_i + \omega_j) \big\} 
\nonumber \\
\exp \left( \pm 2 \pi \mathit{i} \big\{ \frac{(x_2 + x_1)(y_2 - y_1)}{2} \big \} \left( \frac{4}{\sqrt{3} a^{2}} \frac{\phi}{\phi_{0}} \right) \right) + (U \omega_i - \mu) \delta_{ij} \big]
\end{eqnarray}
here $\phi = \frac{\sqrt{3} a^{2}}{4} \bf{B}$, is an uniform magnetic flux in each triangle, $\phi_{0} = \frac{hc}{e}$, is Dirac flux quanta and $(x_1, y_1)$ and $(x_2, y_2)$ are coordinates of a bond on a triangle (see \Fig{fig:triang_lat}). Choosing $\frac{\phi}{\phi_0} = \alpha \in [0,1]$ and $a$ as unity, $h$ reduces as, $h_{ij} = \big[ \big\{ -t_{ij} + t_{ij}^{\prime} (\omega_i + \omega_j) \big\} \exp\left( \pm 2 \pi \mathit{i} \big\{ \frac{(x_2 + x_1)(y_2 - y_1)}{2} \big\} \left( \frac{4}{\sqrt{3}}\right)   \right) \alpha + (U \omega_i - \mu) \delta_{ij} \big]$. This choice of gauge ensures hopping in $x$-direction is $\big[ -\, t_{ij} + t^{\prime}_{ij} \left( \omega_{i} + \omega_{j} \right) \big]$ while hopping in other directions, around a triangle, from $x$-axis is $\left[-t_{ij} + t_{ij}^{\prime} \left( \omega_{i} + \, \omega_{j}\right) \right] \exp \left( \pm 2 \pi \mathit{i} \big\{ \frac{(x_2 + x_1)(y_2 - y_1)}{2} \big\} \left( \frac{4}{\sqrt{3}} \right) \alpha \right)$. The choice of gauge field has produced Hofstadter's butterfly structure on the triangular lattice is shown in~\Fig{fig:hfsdtru0}.~\cite{Yong}   

The Hamiltonian $H$ in \Eqn{eqn:Ham02} shows that $f$-electrons act as an external charge potential or annealed disordered background for itinerant $d$-electrons. This external potential of $f$-electrons can be ``annealed'' to find the minimum total internal energy of the system. It is clear that there is inter-link between subsystems of $f$- and $d$-electrons. This inter-link is responsible for the long range ordered configurations and different charge configurations of $f$- electrons in the ground state.

We set the scale of energy with $t_{\langle ij \rangle} = 1$. The value of $\mu$ is chosen such that filling is ${\frac{(n_{f}~ + ~n_{d})}{2}}$ (e.g. $n_{f} + n_{d} = \frac{1}{2}$ is one fourth-filled case and $n_{f} + n_{d} = 1$ is half-filled case etc.), where $n_{f} = \frac{N_{f}}{N}$, $n_{d} = \frac{N_d}{N}$ and $N$ are density of $f$-, $d$- electrons and total number of sites respectively ($N_f$ and $N_d$ are total number of $f$- and $d$-electrons, respectively). For a lattice of comprising $N$ sites the $H(\{\omega\})$ (given in \Eqn{eqn:Ham02}) is a $N\times N$ matrix for a fixed configuration $\{\omega\}$. For one particular value of $n_{f}$, we choose configuration $\{\omega\} = \{{\omega_{1}, \omega_{2},\cdots, \omega_{N}}\}$. Choosing the parameters $t^{\prime}$, $U$ and $\alpha$, the eigenvalues $\lambda_{i}$($i = 1\cdots N$) of $h(\{\omega\})$ are calculated using the numerical diagonalization technique on the triangular lattice of finite size $N(=L^{2}, L = 12)$ with periodic boundary conditions (PBC). For few cases results are also checked for larger system size (e.g. $L = 24$).  

Our aim is to find the unique ground state configuration (state with minimum total internal energy) of $f$- electrons out of exponentially large possible configurations for a chosen density of $f$-electrons $n_{f}$. In order to achieve this goal, we have used classical Monte Carlo simulation algorithm by annealing the static classical variables $\{\omega\}$ ramping the temperature down from a high value to a very low value. Details of the method can be found here\cite{umesh1,umesh2,umesh3,umesh4,umesh5}.

\begin{figure}
\begin{center}
\includegraphics[width=9.0cm]{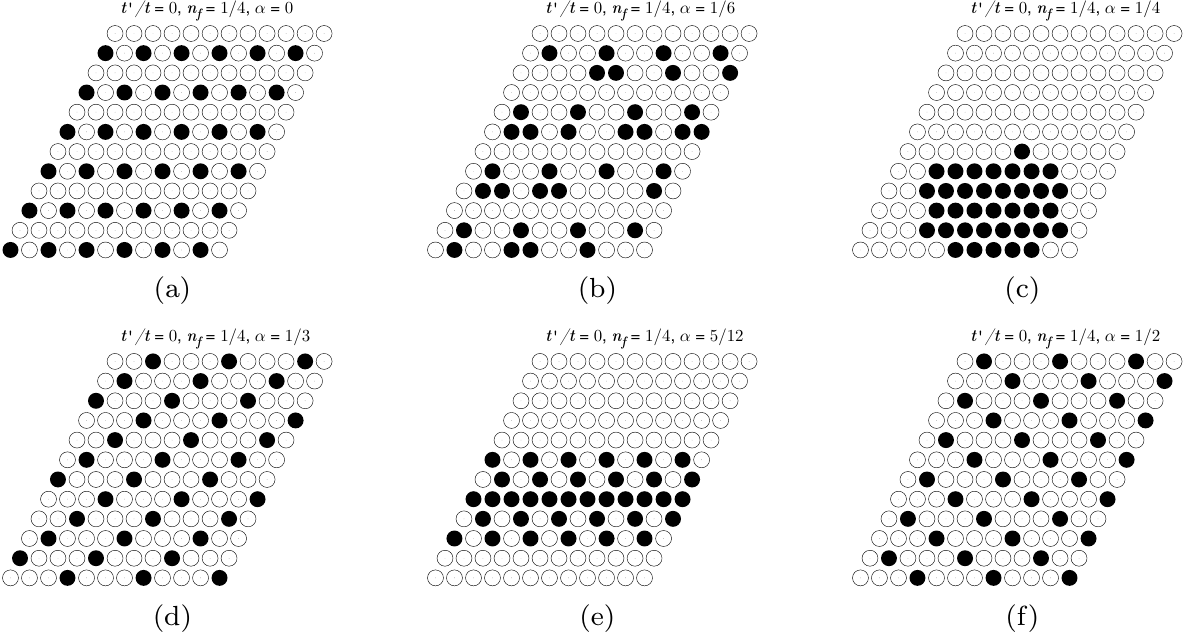}
\caption{Ground state configurations of localized $f$-electrons for $U/t = 1, t^{\prime}/t = 0, n_f = 1/4$ and at (a) $\alpha = 0$, (b) $\alpha = 1/6$, (c) $\alpha = 1/4$, (d) $\alpha = 1/3$, (e) $\alpha = 5/12$ and (f) $\alpha = 1/2$. Here and onwards, in all figures filled circles correspond to the sites occupied by $f$-electrons while open circles correspond to the unoccupied sites.}
\label{fig:conf_tp0nf36}
\end{center}
\end{figure}
\begin{figure}
\begin{center}
\includegraphics[width=9.0cm]{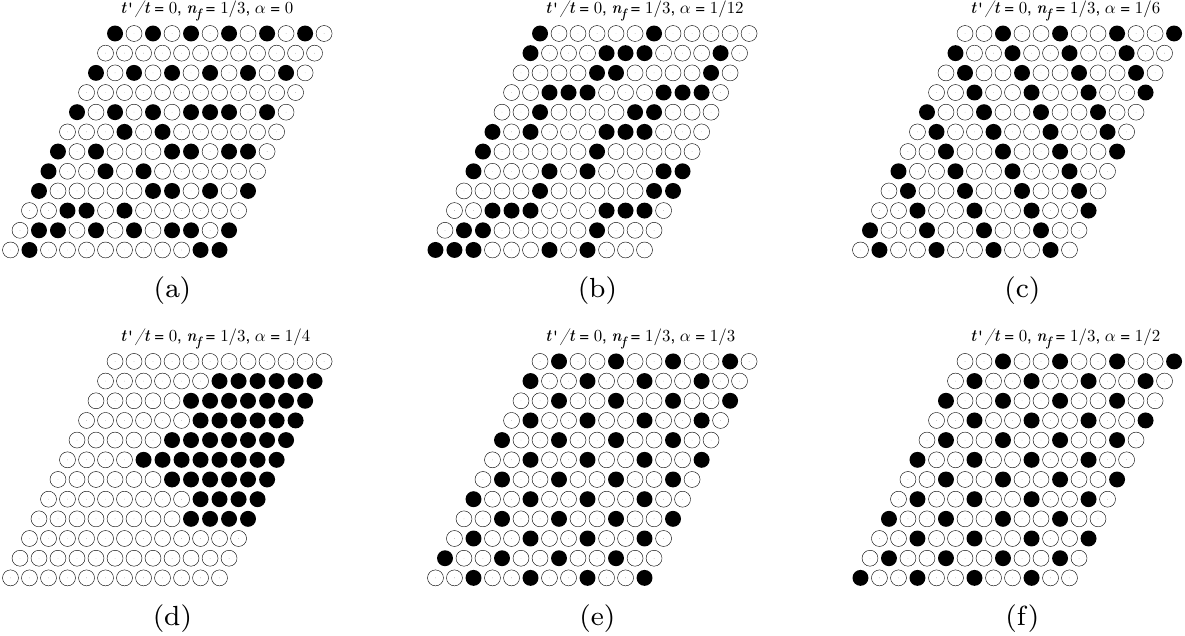}
\caption{Ground state configurations of localized $f$-electrons for $U/t = 1, t^{\prime}/t = 0, n_f = 1/3$ and at (a) $\alpha = 0$, (b) $\alpha = 1/12$, (c) $\alpha = 1/6$, (d) $\alpha = 1/4$, (e) $\alpha = 1/3$ and (f) $\alpha = 1/2$.}
\label{fig:conf_tp0nf48}
\end{center}
\end{figure}
\begin{figure}
\begin{center}
\includegraphics[width=9.0cm]{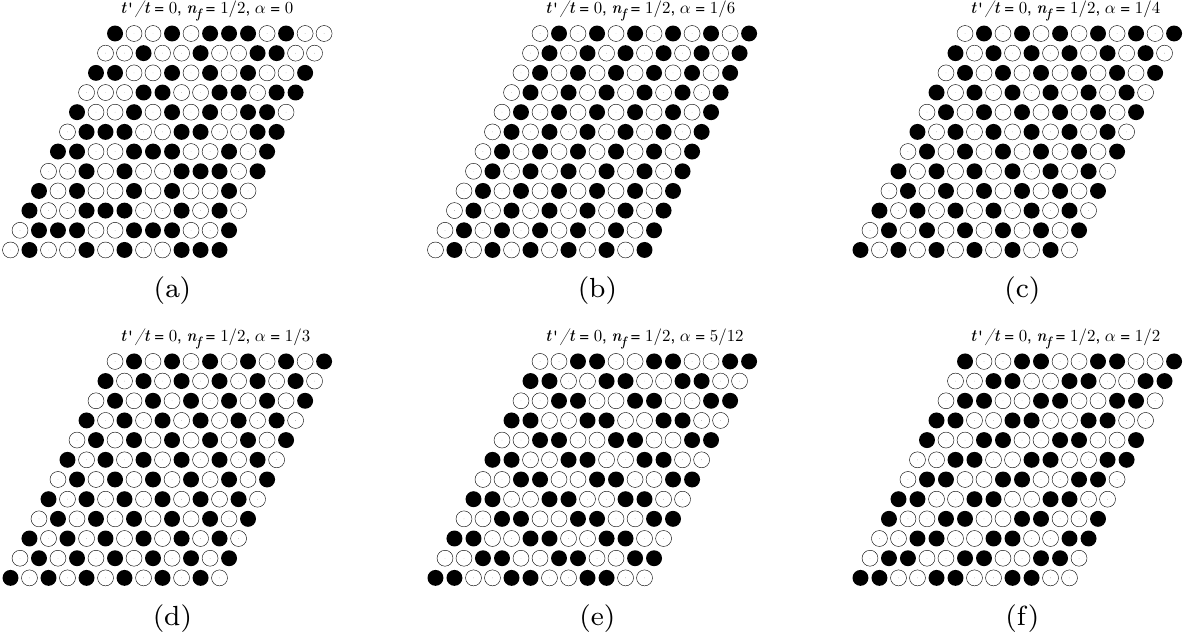}
\caption{Ground state configurations of localized $f$-electrons for $U/t = 1, t^{\prime}/t = 0, n_f = 1/2$ and at (a) $\alpha = 0$, (b) $\alpha = 1/6$, (c) $\alpha = 1/4$, (d) $\alpha = 1/3$, (e) $\alpha = 5/12$ and (f) $\alpha = 1/2$.}
\label{fig:conf_tp0nf72}
\end{center}
\end{figure}
\begin{figure}
\begin{center}
\includegraphics[width=9.0cm]{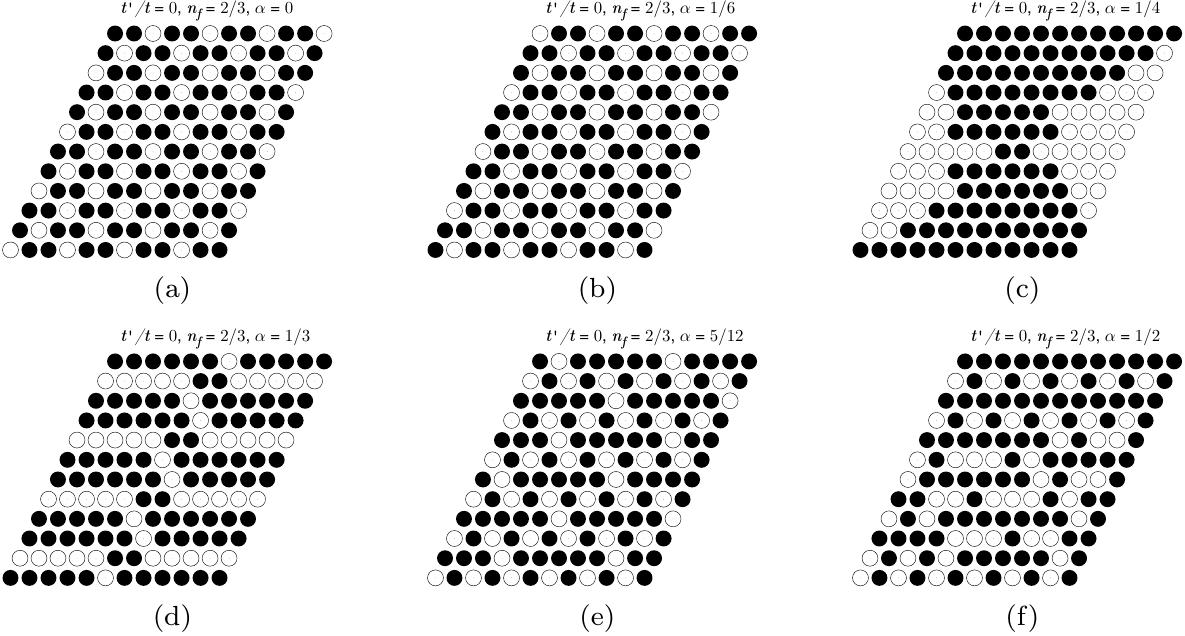}
\caption{Ground state configurations of localized $f$-electrons for $U/t = 1, t^{\prime}/t = 0, n_f = 2/3$ and at (a) $\alpha = 0$, (b) $\alpha = 1/6$, (c) $\alpha = 1/4$, (d) $\alpha = 1/3$, (e) $\alpha = 5/12$ and (f) $\alpha = 1/2$.}
\label{fig:conf_tp0nf96}
\end{center}
\end{figure}

\section{Results and discussion}

In the present paper we have studied the ground state properties of the FKM on a finite strip of triangular lattice with finite magnetic flux $\alpha$ in each triangle for various range parameters like correlated hopping ($t^{\prime}$), on-site interactions between $d$- and $f$- electrons ($U$) and density of $f$-electrons ($n_f$). The value of $U$ is chosen equal to hopping amplitude $t$, where, one would expect prominent role of magnetic flux on the ground state properties. Half filled case i.e. $n_{f} +  n_{d} = 1$ is considered throughout the study. 

Before going to details of the results obtained for various parameters, first let us discuss general nomenclature (for details see~\cite{umesh1}) used further for different types of ground state configurations. Various ground state configurations encountered are mainly: {\it Regular phase}: where the filled sites are arranged in a long range ordered pattern (\Fig{fig:conf_tp0nf36}(a)). There are also situations where a nearly regular pattern with few defects appear. These have been named as `quasi-regular' phase (\Fig{fig:conf_tp0nf48}(a)). {\it Bounded phase}: where several domains of empty sites surrounded by filled sites on all sides (\Fig{fig:conf_tp0nf72}(a)). {\it Hexagonal phase}: where $f$-electrons form hexagonal structures (\Fig{fig:conf_tp0nf96}(a)). {\it Stripe phase}: where the filled sites form axial or diagonal stripes (\Fig{fig:conf_tp0nf72}(b)-(f)). {\it Segregated phase}: where region of sites occupied by $f$-electrons are segregated from the unoccupied sites (\Fig{fig:conf_tp0nf36}(c)). 

\subsection{Correlated hopping $t^{\prime} = 0$}

\begin{figure}
\begin{center}
\includegraphics[width=9.0cm]{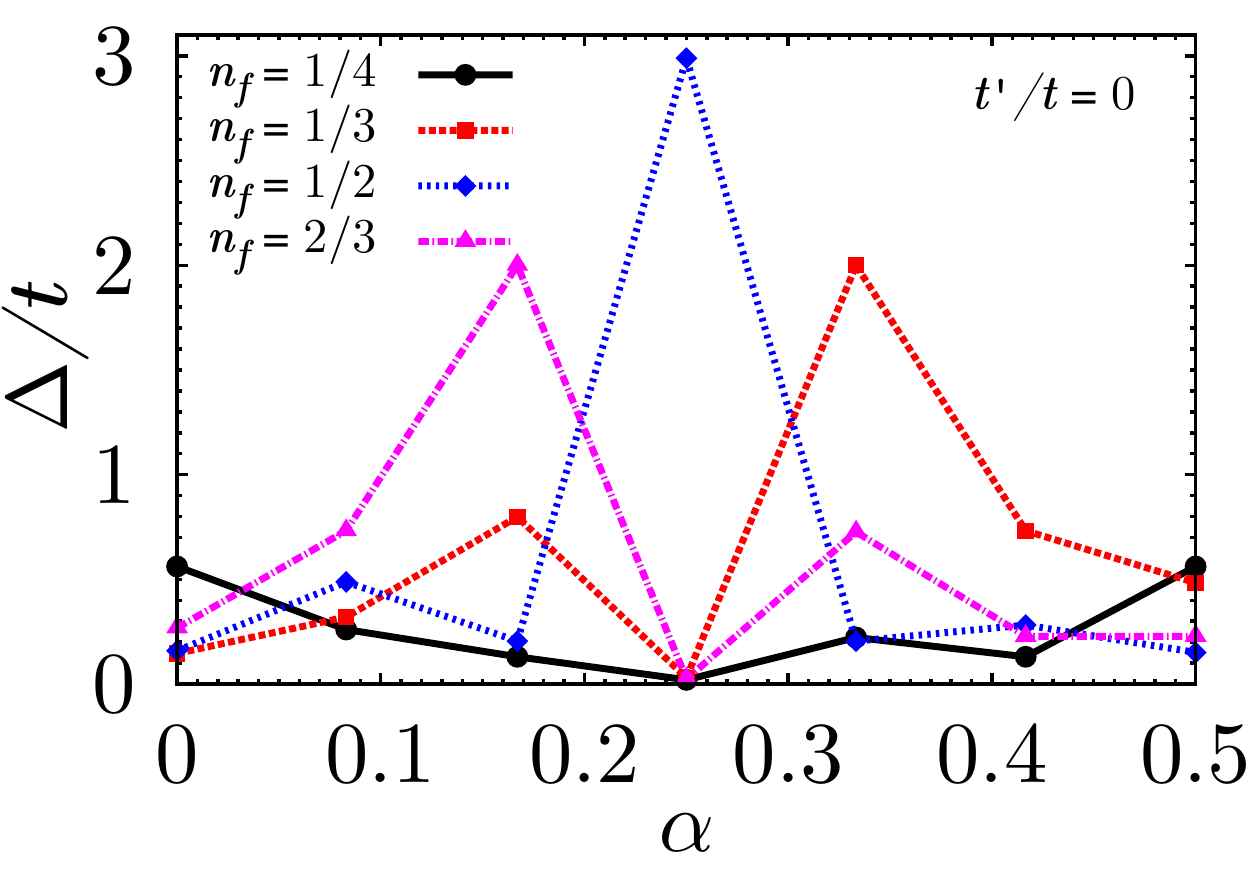}
\end{center}
\caption{(Color online) Flux dependence of gap $\Delta$ at $t^{\prime} = 0$ and for different density of $f$-electrons $n_f$.}
\label{fig:gaptp0}
\end{figure}

First we have shown the effect of magnetic flux on usual FKM ($t^{\prime} = 0$) for $U/t = 1$ and for four different densities of $f$-electrons e.g. $n_f = 1/4$, $1/3$, $1/2$ and $2/3$. 

Ground state configurations of $f$-electrons for $n_f = 1/4$ at different $\alpha$-values are shown in \Fig{fig:conf_tp0nf36}. At different flux values regular, segregated, diagonal and axial stripe patterns of $f$-electrons are observed. It is interesting to note down that at zero flux regular pattern of $f$-electrons was reported (\Fig{fig:conf_tp0nf36}(a))~\cite{umesh1}. As we turn on magnetic field, segregation in $f$-electrons configuration is seen (\Fig{fig:conf_tp0nf36}(b)). At $\alpha = 1/4$, a fully segregated phase is seen (\Fig{fig:conf_tp0nf36}(c)). With further increase in flux different kinds of ordered regular patterns are seen (\Fig{fig:conf_tp0nf36}(d)-(e)). It can be clearly seen that ground state configuration strongly depends on the magnetic field.   

Configurations of minimum energy of $f$-electrons for $n_f = 1/3$ at different $\alpha$-values are shown in \Fig{fig:conf_tp0nf48}. 
At $\alpha = 0$ quasi-regular phase is seen (\Fig{fig:conf_tp0nf48}(a)). With finite $\alpha$, first different kinds of ordered regular configurations of $f$-electrons are observed (\Fig{fig:conf_tp0nf48}(b)-(c)). Again at $\alpha = 1/4$, a fully segregated phase is seen (\Fig{fig:conf_tp0nf48}(d)). With further increase in flux regular ordered patterns are seen (\Fig{fig:conf_tp0nf48}(e)-(f)). At $n_f = 1/3$ also ground state configurations change with change in magnetic field.

\Fig{fig:conf_tp0nf72} shows the configurations of $f$-electrons for density $n_f = 1/2$ and for different $\alpha$-values. At $\alpha = 0$ bounded phase of $f$-electrons (\Fig{fig:conf_tp0nf72}(a)) is seen. Here few unoccupied sites are surrounded by occupied sites by $f$-electrons. With finite flux, axial and diagonal stripe patterns are seen (\Fig{fig:conf_tp0nf72}(b)-(f)). It is observed that flux promotes the ordered phases as at flux $\alpha = 0$ nearly ordered bounded phase was found~\cite{umesh1}. Interestingly at density $n_f = 1/2$ no segregation in configuration of $f$-electrons is seen. Around $\alpha \sim 1/2$ diagonal stripe pattern of $f$-electrons is seen.

Ground state configurations of $f$-electrons for $n_f = 2/3$ at different $\alpha$-values are shown in \Fig{fig:conf_tp0nf96}. At small values of $\alpha$ hexagonal phase is seen (\Fig{fig:conf_tp0nf96}(a)-(b)). A fully segregated phase of $f$-electrons again seen at this filling (\Fig{fig:conf_tp0nf96}(c)). With further increase in magnetic field a regular pattern of $f$-electrons is seen (\Fig{fig:conf_tp0nf96} (d)-(f)).     

In order to see the role of magnetic field on electronic properties of ground state configurations we have studied the density of states (DOS) for various $\alpha$-values (not shown here). A collapse in the energy levels in several groups is seen at finite magnetic flux. Further, a finite gap around Fermi level (energy level corresponding to $N_d$) for ordered regular patterns of $f$-electrons is found. In case of phase segregation infinitesimally small or vanishing gap in DOS is seen around the Fermi level. These results clearly show that magnetic field produces a metal to insulator transition even at finite value of $U$.

Variation of gap $\Delta$ ($ = E(N_{d}+1, N_{f}) +E(N_{d}-1, N_{f})-2 E(N_{d}, N_{f})$) with magnetic flux $\alpha$ at different $n_f$ is shown in \Fig{fig:gaptp0}. At all chosen density of $f$-electrons, $\Delta$ shows irregular dependence on magnetic flux $\alpha$. Except at density $n_f = 1/2$, $\Delta$ goes to zero for $\alpha = 1/4$. It is already noted that $\Delta$ is finite for regular pattern and it tends to zero for segregated phase of $f$-electrons. These results can be understood in terms of the ground state configurations. In case of regular phase, localized $f$-electrons are distributed on throughout lattice. Therefore, itinerant $d$-electrons traveling through the lattice will face the finite on-site interaction on the sites where $f$-electrons are already present. In this case system behaves like an insulator with a finite gap around Fermi-level. While in case of phase segregation, traveling $d$-electrons, find empty space on the lattice where they can travel through without any on-site interaction from $f$-electrons. In this case system behaves like a metal with vanishing gap around Fermi-level.       

These results clearly show that the competition between band energy of $d$-electrons, on-site interaction $U$ and magnetic flux $\alpha$ produces an exotic feature -- \textit{metal to insulator transition} accompanied by segregated phase to an ordered regular phase 
of $f$-electrons.

As we already mentioned that in earlier study several interesting ground state configuarions of $f$-electrons with correlated hopping ($t^{\prime}$), where hopping of $d$-electrons depends on occuapation of $f$-electrons on nearest-neighboring sites, were reported~\cite{umesh1}. Now it will be interesting to see if magnetic field can produce new states in the presence of correlated hopping $t^{\prime}$. Therefore, we have studied the ground state properties of FKM on a triangular lattice with finite magnetic flux $\alpha$ for two values of correlated hopping $t^{\prime} = -1$ and $1$ respectively.  

\subsection{Correlated hopping $t^{\prime} = -1 $}
\begin{figure}
\begin{center}
\includegraphics[width=9.0cm]{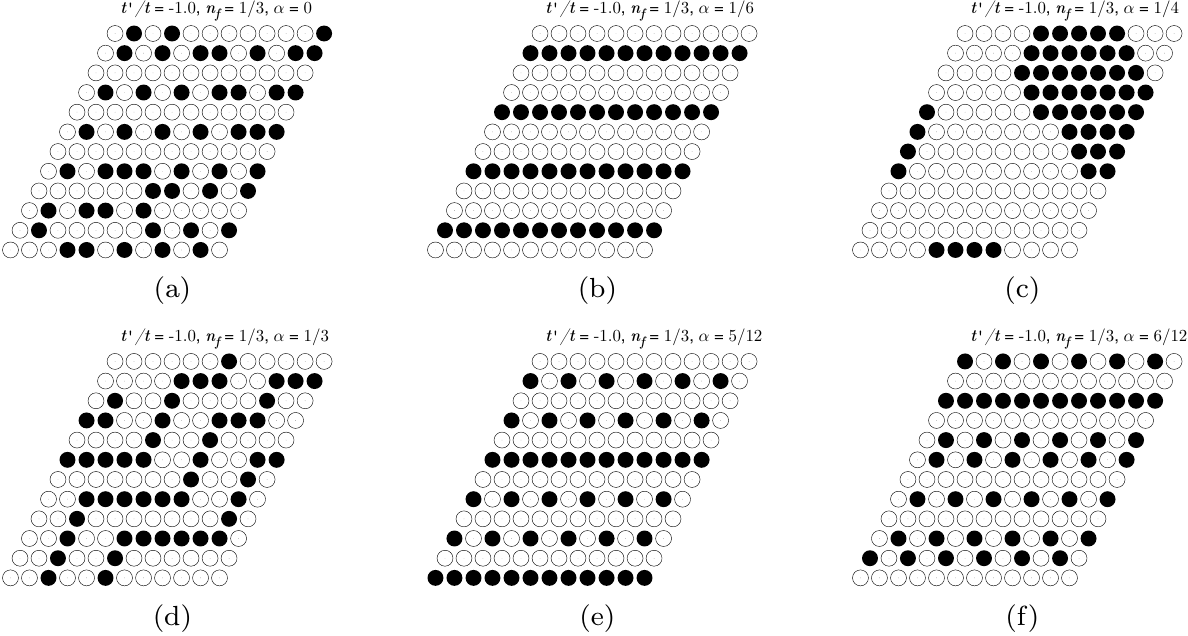}
\caption{Ground state configurations of localized $f$-electrons for $U/t = 1, t^{\prime}/t = -1, n_f = 1/3$ and at (a) $\alpha = 0$, (b) $\alpha = 1/6$, (c) $\alpha = 1/4$, (d) $\alpha = 1/3$, (e) $\alpha = 5/12$ and (f) $\alpha = 1/2$.}
\label{fig:conf_tpm1nf48}
\end{center}
\end{figure}
\begin{figure}
\begin{center}
\includegraphics[width=9.0cm]{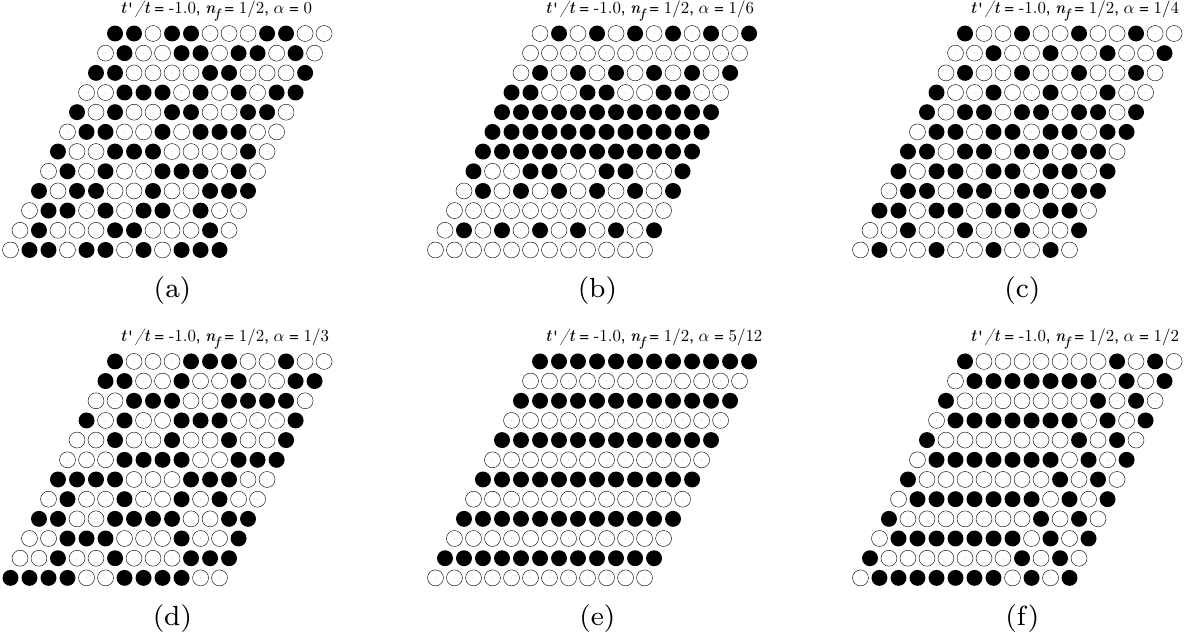}
\caption{Ground state configurations of localized $f$-electrons for $U/t = 1, t^{\prime}/t = -1, n_f = 1/2$ and at (a) $\alpha = 0$, (b) $\alpha = 1/6$, (c) $\alpha = 1/4$, (d) $\alpha = 1/3$, (e) $\alpha = 5/12$ and (f) $\alpha = 1/2$.}
\label{fig:conf_tpm1nf72}
\end{center}
\end{figure}

\Fig{fig:conf_tpm1nf48} shows different ground state configurations of $f$-electrons at density $n_f = 1/3$ for different $\alpha$-values. \Fig{fig:conf_tpm1nf48}(a) shows that ground state configuration is quasi-regular at $\alpha = 0$. Phase segregation takes place with increase in $\alpha$ (\Fig{fig:conf_tpm1nf48}(b)-(c)). With further increase in $\alpha$ values, ground state shows different types of ordered configurations of localized electrons (\Fig{fig:conf_tpm1nf48}(d)-(f)). It seems magnetic field first promotes phase segregation and afterwards ordered regular configurations are promoted.   

\Fig{fig:conf_tpm1nf72} shows the  ground state configurations for different $\alpha$-values at $n_f = 1/2$. Bounded phase of localized electrons was reported earlier at zero magnetic field (\Fig{fig:conf_tpm1nf72}(b)). At $\alpha = 1/6$ a mixture of regular and segregated phase is seen (\Fig{fig:conf_tpm1nf72}(b)). For this density at $\alpha = 1/4$ nearly hexagonal phase is seen (\Fig{fig:conf_tpm1nf72}(c)). At $\alpha = 1/3$ again bounded phase is found (\Fig{fig:conf_tpm1nf72}(d)). With further increase in $\alpha$ axial stripes patterns are identified (\Fig{fig:conf_tpm1nf72}(e)-(f)). Interestingly no phase segregation is again seen at $\alpha = 1/4$ for density $n_f = 1/2$ of $f$-electrons.

We have also studied the ground state configurations of $f$-electrons at $n_f = 1/4$ and $2/3$ for different $\alpha$-values (not show here). In the case of $n_f = 1/4$, for small $\alpha$-values regular arrangement of $f$-electrons in the ground state is seen. With increase in $\alpha$ segregation of localized electrons is found. At $\alpha = 1/4$ fully segregated phase is seen. With further increase in flux an ordered regular patterns are seen. For $n_f = 2/3$, bounded phase is seen at small values of $\alpha$ and at $\alpha = 1/3$. Fully segregated phases are seen at all other chosen values of $\alpha$. It shows that at $n_f = 2/3$ magnetic field strongly promotes the phase segregation at $t^{\prime} = -1$. Further, these results also indicate that the ground state configuration depends on magnetic field at $t^{\prime} = -1$.  

We have also studied the density of states (DOS) with magnetic fields for $t^{\prime} = -1$ and at different densities of $f$-electrons (not shown here). We have found metallic nature (vanishing gap around Fermi level) for segregated and mixture of segregated and hexagonal phases. Insulating nature is found for ordered phases like axial stripes and hexagonal phases. At $t^{\prime} = -1$ again metal to insulator transition accompanied by phase segregation to an ordered phase at finite magnetic field is seen.
\begin{figure}
\begin{center}
\includegraphics[width=9.0cm]{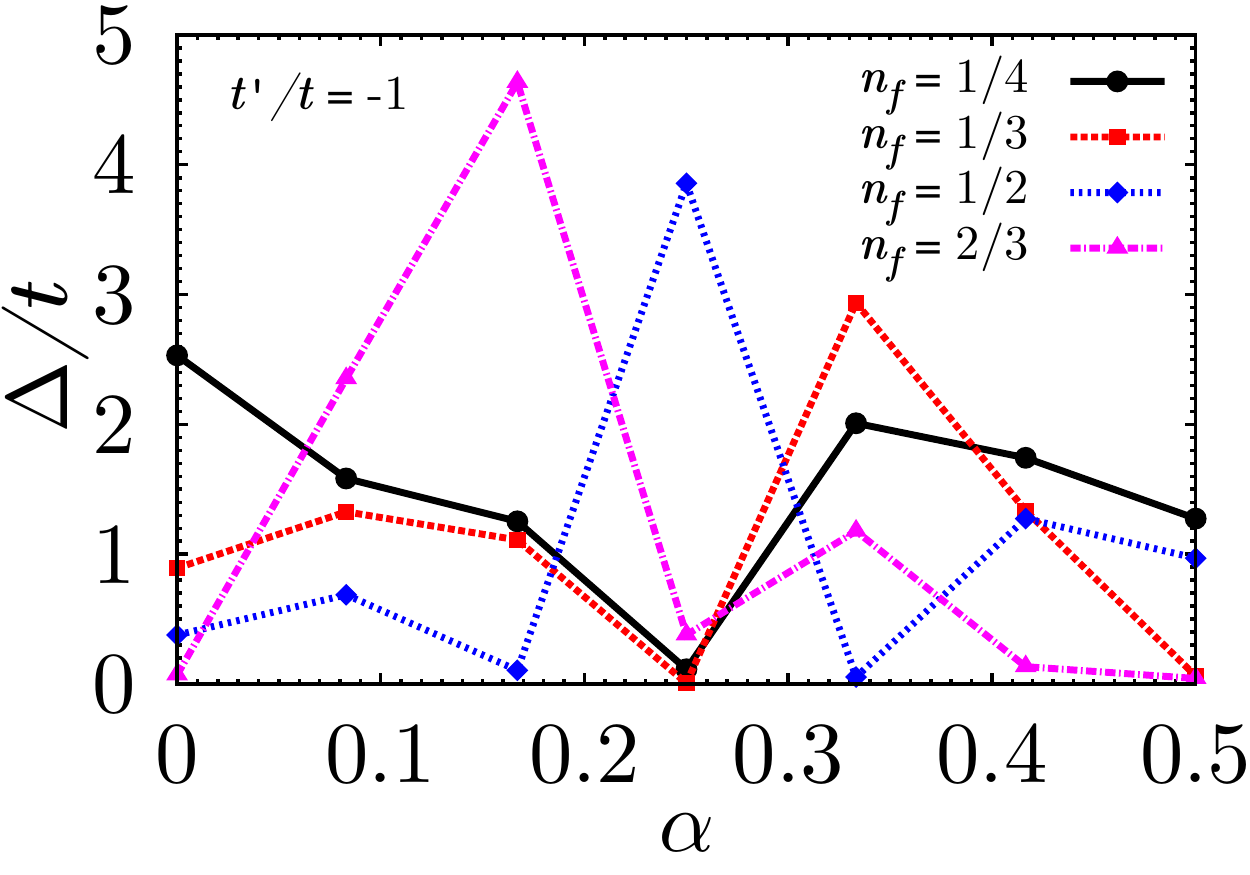}
\end{center}
\caption{(Color online) Variation of gap $\Delta$ with flux $\alpha$ at $t^{\prime} = -1$ and for different density of $f$-electrons $n_f$.}
\label{fig:gaptpm1}
\end{figure}
Variation of gap $\Delta$ with magnetic flux at different $n_f$ is shown in \Fig{fig:gaptpm1}. Similar to the case of $t^{\prime} = 0$, $\Delta$ shows no regular dependence on magnetic field. Similar to the case of $t^{\prime} =0$ (where gap was closing at       $\alpha = 1/4$), here, in addition to $\alpha = 1/4$, $\Delta$ closes at other $\alpha$-values also. These results can be understood by looking at the ground state configurations as discussed earlier.

Again, these results show the competition between band energy of $d$-electrons, on-site interaction $U$ and magnetic flux $\alpha$. The competition between different energies gives rise to \textit{metal to insulator transition} accompanied by segregated phase to an ordered regular phase of $f$-electrons.

\subsection{Correlated hopping $t^{\prime} = 1 $}

\begin{figure}
\begin{center}
\includegraphics[width=9.0cm,height=2.75cm]{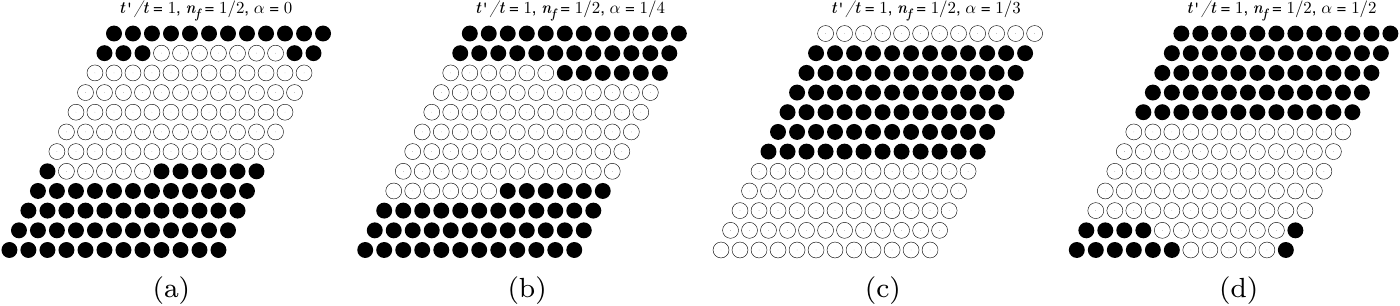}
\caption{Ground state configurations of localized $f$-electrons for $U/t = 1, t^{\prime}/t = 1, n_f = 1/2$ and at (a) $\alpha = 0$, (b) $\alpha = 1/4$, (c) $\alpha = 1/3$ and (d) $\alpha = 1/2$.}
\label{fig:conf_tp1nf72}
\end{center}
\end{figure}

A complete phase segreagtion of localized electrons was earlier reported for correlated hopping $t^{\prime} = 1$~\cite{umesh1}. In order to see the effects of magnetic field on the ground state properties at $t^{\prime} = 1$ we have studied FKM for different densities of $f$-electrons. The ground state configurations of $f$-electrons do not depend on magnetic flux $\alpha$ for all considered values of $n_f$ (only $n_f = 1/2$ case for different values of $\alpha$ is shown in \Fig{fig:conf_tp1nf72}). At all chosen values of $\alpha$ segregated phase is found as ground state configuration of $f$-electrons. It shows that for $U \sim t$, energy gained by the itinerant electrons in phase segregation is more than the potential raised by magnetic flux and on-site interactions. In this case no metal to insulator transition is observed with finite magnetic flux $\alpha$. 

We have also studied the FKM for large values of $U$ (say $U/t = 5$) and at different values of $\alpha$. We have found an ordered regular ground state configuartion with finite gap in the energy spectrum at all chosen values of $\alpha$. These results are in agreement to earlier findings~\cite{taraph02}. At this value of $U$, flux does not have any role on the ground state configurations. 

\section{Summary and conclusion}

In summary, the ground state properties of spinless extended Falicov-Kimball model (FKM) on a triangular lattice with finite magnetic flux perpendicular to the lattice is studied for half-filled case. It is seen that ground state configurations of localized electrons strongly depend on the magnetic flux. Magnetic field produces a metal to insulator transition accompanied by segregated phase to an ordered phase of localized electrons for correlated hopping $t^{\prime} = -1$ and $0$. Ground state configurations have no magnetic field dependence at $t^{\prime} = 1$. These results are applicable to the systems of recent interest like $GdI_2$, $NaTiO_2$ and $MgV_{2}O_4$ etc and can also be seen in the lab for cold atomic systems.  



\begin{thebibliography}{}
\bibitem{orbital} In the presence of finite magnetic field, perpendicular to the plane of triangular lattice, field couples to the spinless itinerant electrons only through canonical conjugate momentum. Due to spinless case under consideration, there is no Zeeman splitting and hence magnetic field couples only with orbital degree of freedom of electrons.
\end{thebibliography}
\end{document}